\documentclass[11pt]{article}
\usepackage{amsmath,amssymb,eps
fig}

\newcommand{\remove}[1]{}

\renewcommand{\phi}{\varphi}

\begin{document}

\title {Theoretical analysis of Crankback  }
\author { Sander Stepanov
\thanks{S. Stepanov  is with the Department of Electrical
Engineering, Technion-Israel Institute of Technology (e-mail:
{sanders@ee.technion.ac.il)}.} \and  Ofer Hadar
\thanks{O.Hadar is with the Department of Electrical
Engineering Systems, Beer-Sheva University, Beer-Sheva,  Israel,
(e-mail:  hadar@bgumail.bgu.ac.il} \and ------
\thanks{------- is with the Department of Electrical
Engineering Systems, Beer-Sheva University, Beer-Sheva,  Israel,
(e-mail:  ----------------}}
\date { }
 \maketitle

\begin {abstract}
asdfafds
\end {abstract}
\bigskip

\section {\textbf{Introduction}} \label {sec1}

Let's designate:
 1. as $x_i$ the simultaneous realization of delay
 at node $i$;
 2. M - means of delays;
 3. V - variance of delay;
 4. T - permitted time to marsh until end node number $n$;
 5. $Q(x,M,V)$ is probability to be more than x for normal;
 distributed variable for means M and variance V, $M >> 0, V << M$;
 6. DistWaste - the distance which the package over to get the end
 node;
 7. f(x, M, V) normal pdf of x for M and V;
 8. $P_{suc}$ - the probability to get end node for old system
 (the criteria is rest time, for example if the rest time is for
 $10 [sec] < 12[sec]=T_{tr}  $(we used 20 sec when it was
 permmitted only $18  = T-T_{tr}=30 - 12  $
 ), when T=30 sec then let's came back;
 9. $P_{suc}^*$ - is  the probability to get end node for new
 system for some parameter $P_{tr}$
 10. $P^*_{SucNeeded}$ is  the probability to get end node for new
 system which we are ready to set for using $P^*_{SucNeeded} < P_{suc}$
\section{The equations for probability to return from node  }
\begin{equation}\label{eq:1}
    P_i=(P (t_i) | NonReturnOnStep_{i-1})
\end{equation}
where $P (t_i)$ the probability to return at node $t_i$
\begin{equation}\label{eq:2}
    P_1=1-P_0(x)=1-\int^{T-[Q_1^*(P_{tr})]^{-1}}_0 f
    (x,M,V)dx=1-F(T-Q_1^*(P_{tr},M,V)
\end{equation}

\begin{equation}\label{eq:3}
    P_2=\int_0 ^{T-[Q_1^*(P_{tr})]} f (x,M,V)
    (1-F(T-x-[Q_2^*(P_{tr})], M,V)dx
\end{equation}

$$
    P_3=\int_0 ^{T-[Q_1^*(P_{tr})]} f (x_1,M,V)
    \int_0 ^{T-[Q_1^*(P_{tr})]^{-1}-x_1}f(x_2, M,V)
    $$
\begin{equation}\label{eq:4}
    (1-F(T-x_1-x_2-[Q_3^*(P_{tr})]^{-1}M,V))dx_1 dx_2
\end{equation}

$$
    P_4=\int_0 ^{T-Q_1^*(P_{tr})}f(x_1,M,V)
    \int_0 ^{T-Q_2^*(P_{tr})-x_1}f(x_2,M,V)
    \int_0 ^{T-Q_3^*(P_{tr})-x_1-x_2}f(x_3,M,V)
    $$
\begin{equation}\label{eq:5}
    (1-F(T-x_1-x_2-x_3-Q_4^*(P_{tr},M,V))dx_1 dx_2 dx_3
\end{equation}

$$
    P_n=\int_0 ^{T-Q_1^*(P_{tr})}f(x_1,M,V)
    \int_0 ^{T-Q_2^*(P_{tr})-x_1}f(x_2,M,V)\ldots
    $$
\begin{equation}\label{eq:6}
    \int_0 ^{T-Q_{n-1}^*(P_{tr})-\sum^{n-2}_{i=1}x_i}
    f(x_{n-1},M,V)(1-F(T-\sum^{n-2}_{i=1}x_i -Q^*_n(P_{tr}),M,V)
    dx_1 dx_2 \ldots dx_{n-1}
\end{equation}
where
\begin{equation}\label{eq:7}
    Q_j^*(P_{tr},M,V)=\{X|(Q(x,(n-j)M,(n-j)V)=P_{tr})\}
\end{equation}

\begin{equation}\label{eq:8}
   P_k=\{P(t_k)|(P(t_{k-1}<T)\}
\end{equation}

    where
    \begin{equation}\label{eq:81}
    P(t_k)=(Q(\sum^{i=n}_{i=k+1}\mu_1,
    \sum^{i=n}_{i=k+1}\sigma^2, T-t_k)>P_{tr})=(Q(\mu(j-k),
    \sigma^2(n-k), T-t_k)>P_{tr}).
    \end{equation}
    For simple approximation $P_i^* \approx  P_i$ can be use difference between coarse
    estimation of probability  to stop the moving at the node i
    \begin{equation}\label{eq:9}
    P_i^{\sum}=(P[T-t_i>Q_i^*(P_{tr},
    M(t_i),V(t_i))]=P[t_i<T-Q_i^*(P_{tr},M(t_i),V(t_i))])=
\end{equation}
$$=Q(T-Q_i^*(P_{tr}, M, V), M*i, V*i)$$
 where
 $M(t_i)=M*t;
V(t_i)=V*i$

\begin{equation}\label{eq:10}
   Q_i^*(P_{tr}, M, V)=\{X \mid Q(X,(n-i)*M,(n-i)*V)=P_{tr}\}
\end{equation}
then
\begin{equation}\label{eq:11}
 P_i^* \approx  P_i^{\sum}-P_{i-1}^{\sum}
 \end{equation}
 For example for T=16 , M=3, V = 1, Ptr = 0.9 where calculated $P_3^*$ = 0.11 and
 $P_5^*$ = 0.08.

    \section{The results of simulations and calculation  }

T=16 , M=3, V = 1, Ptr = 0.9 calculation P1 = 0.193, P2 = 0.194,P3
= 0.138, P4= 0.103; simulation one run P1=0.21, P2=0.194,P3=0.128,
P4=0.103, P5=0.073, P6=0.288;

T=15 , M=3, V = 1, Ptr = 0.9 calculation P1 = 0.553, P2 = 0.159
,P3 = 0.081, P4= 0.052; simulation one run P1=0.54,
P2=0.15,P3=0.085, P4=0.054, P5=0.039, P6=0.125;

T=14 , M=3, V = 1, Ptr = 0.9 calculation P1 = 0.872, P2 =0.056 ,P3
= 0.023, P4=0.014 ; simulation one run P1=0.848,
P2=0.076,P3=0.021, P4=0.018, P5=0.01, P6=0.026;

\section{The equations for optimization  }

$$
    P^*_{suc}=1-P_1-P_2-...-P_n=H(P_{tr})
    $$
Let's designate
$$
    H(P_{tr})=1-P_1-P_2-...-P_n
    $$
then
    $$
    P_{tr}=H^{-1}(P^*_{suc})=H^{-1}(P_{suc}*k) ,k<1
$$

\begin{equation}\label{eq:12}
    P_{tr Opt}=\arg \min_{P_{tr}} | H(P_{tr})-P^*_{SucNeeded} |
\end{equation}
\section{The equations for means waste distance  calculation   }
\begin{equation}\label{eq:13}
    P_1+P_2+P_3+...+P_{n-1}+P_{suc}^*=1.0
\end{equation}

\begin{equation}\label{eq:14}
    P_k^*=[\prod^{i=k-1}_{i=1}(1-P_{suc}^*)]P_{suc}^*
\end{equation}
where k is the number  of efforts  to get the end node and $P_k^*$
is       the probability to get for end node for k attempts
\begin{equation}\label{eq:15}
    \widehat{dist}_{waste}=(\sum^{i=n-1}_{i=1}P_i * 2* t*i)
\end{equation}
where the t is the distance between nods, if the distances are
equal, if not equal it is possible to use the average distance.
Then the average time in travel to rich the end not is
\begin{equation}\label{eq:16}
    M(DistWaste)=\sum^{k=\infty}_{k=1}P_k^*\widehat{dist}_{waste}*k]+t*n
\end{equation}
or it is by another version
\begin{equation}\label{eq:17}
    M(DistWaste)=(\sum^{k=n-1}_{k=1}P_k*2*M*k])/P_{suc}^*
\end{equation}
for example let's was 1000 attempts for 4 nodes by this way: 100
times  - success ( was rich node N3), 200 times - was return from
node N1 (each time was waste 2*M [sec]) , 700 times was return
form node N2 (each time was waste 2*M*2 [sec] than the average
waste for one success is (200*2*M + 700 * 2 * M * 2)/100 let's
divide the divider and divided by number of efforts (in our case
it is 1000) now we get ((200*2*M + 700 * 2 * M *
2)/1000)/(100/1000) or $ (P_1*2*M + P_2*2*M*2)/P_{suc}^*$

\end{document}